\documentstyle[12pt]{article}
\title{\Large\bf QUANTIZATION OF A $q$-DEFORMED\\
FREE RELATIVISTIC PARTICLE}

\author{R.P.Malik
\thanks{E-mail:MALIK@THEOR.JINRC.DUBNA.SU}\\
      \small \it Bogoliubov Laboratory of Theoretical Physics,\\
 \small \it JINR, 141980 Dubna, Moscow Region, RUSSIA }

\begin{document}
\hoffset=-1truecm
\voffset=-2truecm
\baselineskip=16pt
\date{}
\maketitle
\begin{abstract}
A $q$-deformed free scalar relativistic particle is
discussed in the framework of the BRST formalism. The $q$-deformed
local gauge symmetry and reparametrization invariance of
the first-order Lagrangian have been exploited for the BRST quantization
of this system on a $GL_{q}(2)$ invariant quantum world-line. The on-shell
equivalence of these BRST charges requires
the deformation parameter to be $ \pm 1 $ under certain
identifications.The same restriction ($ q= \pm 1 $) emerges from
the conservation of the
$q$-deformed BRST charge on an arbitrary (unconstrained) manifold and the
validity of the BRST algebra. The solutions for the equations of motion
respect $GL_{q}(2)$
invariance on the mass-shell at any arbitrary value of the evolution
parameter characterizing the quantum world-line.
\end{abstract}
\pagenumbering{arabic}
\newpage
\noindent
Symmetry groups and symmetry algebras are some of the firm pillars on which
the whole edifice of the modern developments in theoretical physics rests.
Mathematically, the $q$-deformed symmetry algebras
(compact matrix pseudo-groups or quantum groups [1,2])
are examples of the quasi-triangular Hopf algebras [3]. These $q$-deformed
groups and algebras have recently been the subject of considerable interest
in the hope of
developing some more general symmetries that might have profound implications
in very sensitive physical theories (where $q$ is very close to one) [4].
In spite of considerable progress in
the mathematical direction, the
key concepts of quantum groups have not penetrated into the realm of
physical applications in an overwhelming and compelling manner. Some attempts
have been made,
however, to see the impact of these groups in the context of
$q$-deformed gauge
theories [5] as well as in few well-known physical examples [6]. These groups
are also conjectured to provide a  fundamental length in the context of
space-time quantization [7] with a non-commutative underlying geometry
of the space-time manifold [8]. These objects have manifested
themselves in statistical systems, conformal field theories, knot theory,
nuclear physics, etc.[ see, e.g., Ref.9 and references therein].

Recently, a Lagrangian formulation has been developed to describe
a $q$-deformed scalar as well as a spinning
relativistic particle in a consistent and cogent way [10].
In this approach, the Lorentz invariance is respected throughout
the discussion, which might turn out to be useful in the development
of the Lorentz covariant $q$-deformed field theories. The main objective
of the present paper is
to develop the BRST formalism for the $q$-deformed scalar particle
of Ref. [10] on a
$GL_{q}(2)$ invariant quantum world-line defined on a {\it flat} but
{\it q-deformed} cotangent manifold to the Minkowski space-time
(configuration) manifold.
We derive $q$-(anti)commutation relations for this system which are
(graded)associative on the mass-shell and the on-shell. One of
the key features of
our work is the $GL_{q}(2)$ invariance of the solutions for the equations of
motion on the {\it mass-shell} at any arbitrary value of the evolution
parameter. The BRST quantization has been carried out by exploiting the
local gauge symmetry and the reparametrization invariance of the starting
$q$-deformed Lagrangian. The equivalence of the BRST charges
corresponding to these symmetries requires that the deformation parameter
$q$ must be $\pm 1$. This condition ($ q= \pm 1$) also emerges from the
conservation of the BRST charge on the unconstrained manifold and the
requirement that the BRST algebra should be satisfied. We do not discuss here
the $q$-deformed Hamiltonian formulation, $q$-deformed Dirac brackets, etc.,
for the above system. The $q$-deformed Hamiltonian formulation for a
scalar as well as a spinning particle would be reported in a future
publication [11].

We start off with three equivalent Lagrangians for an undeformed (classical)
free relativistic particle [12] moving on a world-line embedded in
a $D$-dimensional flat Minkowski manifold.
The mass-shell condition ($ p^2 - m^2 = 0 $) is a common feature
of the first-order Lagrangian
($ L_{F} = p_{\mu} \dot x^{\mu} - \frac{e}{2} (p^2 - m^2)$), the second-order
Lagrangian ($ L_{S}= \frac{1}{2} e^{-1} \dot x^2 + \frac{1}{2} e m^2$) and
the Lagrangian with a square-root ($L_{0}= m (\dot x^2)^{1/2}$).
Except the mass (cosmological constant)
parameter ($m$), the target space canonically conjugate coordinates ($x^\mu$)
and momenta ($p_{\mu}$) as well as the einbein field ($e$) are functions of
an evolution parameter ($\tau$) characterizing the trajectory of the free
motion of a relativistic particle and $ \dot x^\mu = \frac{d x^\mu}{d\tau}$.
All the above dynamical variables are
the {\it even} elements of a Grassmann algebra.
The first- and second-order Lagrangians
are endowed with first-class constraints
$ \Pi_{e} \approx 0$ and $ p^2 - m^2 \approx 0 $, where $\Pi_{e}$ is
the conjugate momentum corresponding to the einbein field $e$. For the
covariant canonical quantization of such systems, the most suitable approach
is the BRST formalism [13]. The BRST invariant Lagrangian ($L_{brst}$)
corresponding to the first-order Lagrangian ($L_{F}$) is [14]
\begin{equation}
L_{brst} = p_{\mu} \dot x^{\mu} - \frac{e}{2} (p^2 - m^2) + b \;\dot e
+ \frac{b^2}{2}  + \dot {\bar c}\; \dot c,
\label{1}
\end{equation}
where the even element $b$ is the Nakanishi-Lautrup auxiliary field and
(anti)ghost fields $(\bar c)c$ are the {\it odd} elements of a
Grassmann algebra ($ c^2= 0,\; \bar c^2=0 $). In the BRST quantization
procedure, the first-class constraints
$ \Pi_{e}= b \approx 0 $ as well as  $ p^2 - m^2 \approx 0 $ turn up
as constraints on the physical states when one requires that
the conserved and
nilpotent BRST charge $Q_{brst}
= \frac{c}{2}( p^2 - m^2 ) + b \dot c $ must annihilate the
physical states in the
quantum Hilbert space. The conservation of the BRST charge
on any arbitrary unconstrained manifold is ensured by the
equations of motion $ \dot p_{\mu}= 0,\; \dot b = -\frac{1}{2} (p^2-m^2),
\;\ddot c = \ddot {\bar c} = 0,\; b+\dot e = 0,\; \dot x_{\mu} = e p_{\mu} $.

To obtain the $q$-analogue of the above Lagrangian ($L_{brst}$), we follow
the the prescription of Ref. [10] where the configuration space
corresponding to the Minkowski space-time manifold
is {\it flat} and {\it undeformed} ($ x_{\mu} x_{\nu}
= x_{\nu}  x_{\mu} $) but the cotangent
manifold (momentum phase space) is $q$-{\it deformed} ($ x_{\mu} p_{\nu}
= q\; p_{\nu}  x_{\mu},\;
x_{\mu} x_{\nu} = x_{\nu}  x_{\mu},\; p_{\mu}p_{\nu}= p_{\nu} p_{\mu}$)
in such a way that
the Lorentz invariance is preserved for any arbitrary ordering of
$\mu$ and $\nu$. Here all the dynamical variables are taken
as hermitian elements of an algebra in involution ($ |q|= 1 $) and $q$ is a
non-zero $c$-number. As a consequence of the above deformation, the
following {\it on-shell} and {\it (graded)associative } $q$-(anti)commutation
relations emerge
\footnote {These on-shell $q$-(anti)commutation relations
emerge from the basic (un)deformed relations on a deformed cotangent
manifold, the equations of motion obtained from
the (un)deformed BRST-invariant Lagrangians (1) {\it or} (8) and by
exploiting the {\it mass-shell} condition $p^2 - m^2 = 0$.
For instance, it can be readily seen that if we take the {\it on-shell}
conditions only, there is a contradiction between the relations
$\dot b\; p_{\mu}= q\; p_{\mu}\; \dot b$ and $p_{\mu}\;p_{\nu}
= p_{\nu}\;p_{\mu}$
with $\dot b= - \frac{1}{2}( p^2 - m^2 )$. Thus, in the computation of the
$q$-(anti)commutation relations for the BRST invariant Lagrangians,
the {\it mass-shell} as well
as the {\it on-shell} conditions should be exploited together.}
\begin{eqnarray}
&& x_{\mu}\;
x_{\nu}= x_{\nu}\; x_{\mu}, \quad \dot x_{\mu}\; \dot x_{\nu}
= \dot x_{\nu}\;
\dot x_{\mu},\quad \dot x_{\mu} \;x_{\nu}= x_{\nu} \dot x_{\mu}, \quad
x_{\mu} \dot x_{\nu}= \dot x_{\nu} x_{\mu}, \nonumber\\
&&p_{\mu}\;p_{\nu}= p_{\nu}\;p_{\mu},\quad  x_{\mu}\;p_{\nu}
= q\; p_{\nu}\; x_{\mu}, \quad \dot
x_{\mu}\; p_{\nu} = q\; p_{\nu}\;\dot x_{\mu}, \quad e\;x_{\mu} =
q\;x_{\mu} e, \nonumber\\ &&e\;p_{\mu} = q\;p_{\mu}\;e, \quad    e\;\dot
x_{\mu}\; = q\; \dot x_{\mu}\; e , \quad e\;b = b\;e, \quad e\;c = c\;e,
\quad e\;\bar c= \bar c\; e, \nonumber\\
&& c\;\bar c = -\frac{1}{q}\bar c\; c, \qquad c\;\dot {\bar c}
= -\frac{1}{q} \dot {\bar c} \; c,  \qquad
\dot c\;\bar c = -\frac{1}{q} \bar c\; \dot c,
\qquad \dot c\;\dot {\bar c} = -\frac{1}{q} \dot{\bar c}\; \dot c,
\nonumber\\
&&c\;x_{\mu}= q\;x_{\mu}\;c,\quad \bar c\;x_{\mu}= q\;x_{\mu}\;\bar c,
\quad c \;p_{\mu}= q\;p_{\mu}\;c,\qquad
\bar c \;p_{\mu}= q\;p_{\mu}\;\bar c,\nonumber\\
&& b\;x_{\mu}=q \;x_{\mu}\;b,\qquad
b\;p_{\mu}=q \;p_{\mu}\;b,\qquad b\;c\;=c\;b\;,\qquad b\;\bar c\;
=\bar c\;b\;, \nonumber\\
&&e\;m = q\;m\;e, \quad x_{\mu}\; m = q\;m\;x_{\mu}, \qquad
p_{\mu}\;m = m\;p_{\mu},\quad  c\;\dot c = - q\;\dot c\;c, \nonumber\\
&&b\;m = q\; m\;b,\quad c\;m = q \;m\;c, \quad \bar c\;m
= q\;m\;\bar c, \quad c^2 = 0, \quad \bar c^2 = 0,
\label{2}
\end{eqnarray}
where the mass-shell condition $p^2 -m^2=0$, emerging from the equations of
motion $ \dot b= - \ddot e = -\frac{1}{2}(p^2 -m^2)=0 $ has been
{\it imposed}.
Mathematically, this restriction implies that the $b$ field is
$\tau$-independent and the $\tau$-dependence of the einbein field is
at most linear. Physically, it just means that the mass-shell condition
is {\it strongly} equal to zero on the quantum world-line
even in the case of the $q$-deformed BRST formalism.
It is straightforward to see that in the limit when
the  odd  Grassmann variables ($c,\bar c$) and the  even variable
($b$) are set
equal to zero, we obtain the $q$-commutation relations for a $q$-deformed
scalar free relativistic particle of Ref. [10] and in the limit
$ q \rightarrow 1 $ the usual (anti)commutation relations among the
dynamical variables of the Lagrangian (1) emerge automatically.

Before obtaining the $q$-deformed Lagrangian,
it is essential to define a $q$-deformed world-line for the free motion
of a scalar relativistic particle on the cotangent manifold
because the $q$-deformation is
present in this manifold and the Lagrangian has to describe the motion
on this specific quantum world-line.
Such a $GL_{q}(2)$ invariant world-line, consistent with the
$q$-(anti)commutation relations (2), can be
defined in terms of the coordinate generator $x^\mu$ and the momentum
generator $p_\mu$ as [10]
\begin{equation}
x_{\mu}(\tau)\; p^{\mu}(\tau) = \;q\; p_{\mu}(\tau)\; x^{\mu}(\tau),
\label{3}
\end{equation}
where repeated indices are summed over (i.e., $ \mu= 0,1,2........D-1 $),
and the world-line is parameterized by a real commuting variable $\tau$.
The following $GL_{q}(2)$ transformations
\begin{eqnarray}
x_{\mu} \; &\rightarrow&\; A\; x_{\mu} + B\; p_{\mu}, \nonumber\\
p_{\mu} \; &\rightarrow&\; C\; x_{\mu} + D\; p_{\mu},
\label{4}
\end{eqnarray}
are implied in the component pairs:
$(x_{0},p_{0}),(x_{1},p_{1}).........(x_{D-1},p_{D-1})$ of the phase
variables in equation (3) and its form-invariance can be readily
checked if we assume the commutativity
of the phase variables with
elements $A,\;B,\;C$, and $D$  of a $ 2 \times 2 \quad GL_{q}(2)$
matrix obeying the
braiding relationship in  rows and columns as:
\begin{eqnarray}
AB &=& q\;BA, \quad AC = q\;CA, \quad CD = q\; DC,\quad BD=q\;DB,
\nonumber\\
BC &=& CB,\qquad \; \;  AD - DA =\;(q\; -\; q^{-1}) \; BC.
\label{5}
\end{eqnarray}
It will be noticed that there is another candidate, namely;
\begin{equation}
c(\tau)\;\bar c(\tau) = -\frac{1}{q}\;\bar c(\tau)\; c(\tau),\qquad
c^2(\tau)=0, \qquad \bar c^2(\tau) = 0 ,
\label{6}
\end{equation}
which also remains form-invariant under the following transformations
\begin{equation}
\left(\begin{array}{c}
c \\
\bar c\\ \end{array} \right) \quad
\rightarrow
\left(\begin{array}{cc}
A, & B \\
C, & D \\ \end{array} \right) \;
\left(\begin{array}{c}
c \\
\bar c\\ \end{array} \right),
\label{7}
\end{equation}
if we assume the commutativity of the (anti)ghost fields $(\bar c) c$
with the elements
$A, \; B,\; C,\;$ and $D$ of the $GL_{q}(2)$ matrix obeying relations (5).
However, it
cannot be taken as the definition of the quantum world-line because
these fields are totally decoupled from the rest of the
theory and their on-shell conditions
$ \ddot c = \; \ddot {\bar c } =\; 0 $
do not lead to anything interesting and substantial.

The BRST invariant first-order Lagrangian ($ L_{f} $) that describes the free
motion ($\dot p_{\mu}=0$) of a free $q$-deformed relativistic particle is
\begin{equation}
L_{f} = q^{1/2} p_{\mu} \dot x^{\mu} - \; \frac{e}{1+ q^2} (p^2-m^2)
+b \;\dot e + \frac{b^2}{1+q^2}  + \dot {\bar c}\; \dot c,
\label{8}
\end{equation}
where the $q^{1/2}$ factor appears in the first term due to the Legendre
transformation with $q$-symplectic metrices [10]
\begin{equation}
\Omega_{AB}(q)= \left(\begin{array}{cc}
0, & -q^{-1/2} \\
q^{1/2}, & 0 \\ \end{array} \right) \qquad  \mbox{and} \qquad
\Omega^{AB}(q) = \left(\begin{array}{cc}
0, & q^{-1/2} \\
-q^{1/2}, & 0 \\ \end{array} \right).
\label{9}
\end{equation}
In the third term of the Lagrangian (8), there is no $q^{1/2}$ factor
because the canonically conjugate variables $e$ and $b$ commute
($ e\;b = b\;e, \dot e\;b = b\;\dot e $). Therefore, the standard
canonical symplectic metrices (i.e., $ q = 1 $ in equation (9)) have to
be exploited for the Legendre transformations for these fields.
Here the $q$-BRST Hamiltonian for a free relativistic particle
has been taken to be:
$ H = \frac{e}{1+ q^2} (p^2 - m^2)- \frac{b^2}{1+q^2}
+ \dot {\bar c} \dot c $.
The equations of motion from the Lagrangian (8) on the mass-shell are
\begin{eqnarray}
\dot x_{\mu} &=& \;q^{1/2}\;  e \; p_{\mu},\quad
\dot p_{\mu} = \;0, \; \quad \; \ddot c\; =\; 0 ,
\quad \ddot {\bar c } = 0, \nonumber\\
\dot b\;&=&\;- \frac{q^4}{1+q^2}\; ( p^2 - m^2 ) = 0, \qquad
\dot e\;= - \;b,
\label{10}
\end{eqnarray}
which satisfy the {\it on-shell} and the {\it mass-shell}
$q$-(anti)commutation relations (2).
In the derivation of the equations of motion from the Lagrangian (8), the
$GL_{q}(2)$ invariant differential calculus  has been exploited [15].
For instance, for {\it even} dynamical variables obeying $x y = q y x$,
any  monomial in the Lagrangian (8) is arranged in the form $y^r\; x^s$,
and then we use
\begin{eqnarray}
\frac{\partial ( y^r \;  x^s )}{\partial  x}
&=&\; y^{r}\; x^{s-1}\;q^r \; \frac{( 1 - q^{2s})}{( 1 - q^{2})},
\nonumber\\
\frac{\partial ( y^r \;  x^s )}{\partial  y}
&=&\; y^{r-1}\; x^{s}\; \frac{( 1 - q^{2r})}{( 1 - q^{2})},
\label{11}
\end{eqnarray}
where  $ r,s \in {\cal Z} $ are whole numbers (not fractions). For
the differentiations with respect to the odd Grassmann variables $ \dot c\;
\mbox{and} \; \dot {\bar c} $,
these variables are first
brought to the left side in the corresponding expressions by using
$q$-(anti)commutation relations (2)  and, only then, differentiation
is carried out.

It is rather cumbersome to obtain general solutions for the equations of
motion (10) for any arbitrary dependence of the dynamical variables on the
evolution parameter $\tau$.
The mass-shell condition ($ p^2-m^2 = 0 $), however, emerging from the
restriction $ \ddot e = - \dot b = 0 $ comes to our rescue.
The $GL_{q}(2)$
invariant solutions for the equations of motion (10), under such
restriction, are
\begin{eqnarray}
x_{\mu}(\tau)&=& x_{\mu}(0) + q^{1/2} e(0) p_{\mu}(0) \tau
-\frac{1}{2} \; q^{1/2}\;b(0)\; p_{\mu} (0)\;\tau^2 ,\nonumber\\
e\;(\tau)\; &=&  e(0)\; - \;b(0)\; \tau, \nonumber\\
c\;(\tau)\; &=&  c(0)\; + f\;\tau, \nonumber\\
\bar c\;(\tau)\; &=& \bar c(0)\; + \bar f\;\tau, \nonumber\\
b\;(\tau)\; &=& b(0), \nonumber\\
p_{\mu}(\tau)&=& p_{\mu}(0),
\label{12}
\end{eqnarray}
where $f$ and $\bar f$, present in the solutions for $c(\tau)$ and
$\bar c(\tau)$, are $\tau$-independent {\it odd} elements of a Grassmann
algebra ($ f^2 = \bar f^2 = 0 $) and they obey the following
$q$-anticommutation relations with the rest of the  odd
dynamical variables
\begin{eqnarray}
f\;{\bar f} &=& -\frac{1}{q}\; {\bar f}\; f, \quad
f\;{\bar c} = -\frac{1}{q}\; {\bar c}\; f,
\quad f^2\;= \bar f^2 = 0, \nonumber\\
\quad  c\; {\bar f} &=& -\frac{1}{q} {\bar f}\; c, \quad\;
c\; f = - \; f\; c, \quad\; {\bar c}\; {\bar f} = - {\bar f}\; {\bar c}.
\label{13}
\end{eqnarray}
The $q$-commutation relations of $f$ and $\bar f$ with the rest of the
 even dynamical variables are the same as that of  $c$ and
$\bar c$ in equation (2). With equations
(2), (13) and solutions (12), it is interesting to check that
 all these relations and the
$GL_{q}(2)$ invariant quantum world-line (3) are invariant for any arbitrary
value of the evolution parameter $\tau$, if we assume the validity of these
relations at initial "time" $\tau=0$.
The second-order Lagrangian ($L_{s}$), describing the motion of a scalar
relativistic
particle on the tangent manifold (velocity phase space), can be obtained
from the first-order
Lagrangian (8) by exploiting equations (2) and (10) as given below:
\begin{equation}
L_{s} = \; \frac{q^2}{1 + q^2}\; e^{-1} \; ( \dot x_{\mu})^2
+ \; \frac{e}{1 + q^2}\;  m^2
+ b\; \dot e + \frac{b^2}{1+q^2}  + \dot {\bar c}\; \dot c.
\label{14}
\end{equation}
The consistent expression for the
canonical
momenta ($ p_{\mu} $) and the rest of the canonical momenta
($\Pi's$) for the
first- and second-order Lagrangians (8) and (14) are
\begin{eqnarray}
p_{\mu}\; &=& \; q^{-3/2} \Bigl ( \frac{\partial L_{(f,s)}}{\partial
\dot x^{\mu}} \Bigr ) \equiv q^{-1/2}\;e^{-1}\;
\dot x_{\mu} ,  \qquad  \Pi_{e} =\; b,\nonumber\\
\Pi_{b} &=& \;0 , \; \qquad \;  \Pi_{c} = -\;q\;\dot {\bar c}, \; \qquad
\; \; \Pi_{\bar c} = \dot c.
\label{15}
\end{eqnarray}
Due to the $GL_{q}(2)$ invariant differential calculus developed in
Ref. [15], the differentiation of the Lagrangian $L_{s}$ with respect to
the einbein field $e$ yields
\begin{equation}
\dot b = \frac{q^4}{1 + q^2}
\Bigl [ m^2 \; - q^{-1}\;e^{-1}\;(\dot x_{\mu})\; e^{-1}\;(\dot x^{\mu})\;
\Bigr ],
\label{16}
\end{equation}
which is consistent with the corresponding equation of motion
derived from the first-order
Lagrangian ($L_{f}$) and equation (15). In fact, the above second-order
Lagrangian is equivalent to the first-order Lagrangian in all aspects.

It has been demonstrated in Ref. [10] that the first-order Lagrangian for the
$q$-deformed scalar particle ($ L_{\cal F}
= q^{1/2} p_{\mu} \dot x^{\mu} - \; \frac{e}{1+ q^2} (p^2-m^2)$) is endowed
with the $q$-deformed gauge and reparametrization symmetries which are
found to be on-shell equivalent only for $ q = \pm 1$.
For an arbitrary value of $q$, the above symmetries are
{\it not} equivalent. Thus, both of these symmetries can be exploited for
the BRST quantization. For instance, it can be seen that the
Lagrangian (8) is
invariant under the following nilpotent BRST transformations
\begin{eqnarray}
\delta_{B} x^{\mu}&=& q^{1/2}\;\eta\;c\;p^{\mu}, \qquad \delta_{B} c\; = \;0,
\quad \;\; \delta_{B} b\; = 0, \nonumber\\
\delta_{B} p^{\mu} &=& 0, \qquad
\delta_{B} \bar c = q^2 \;\eta\;b \;, \qquad
\delta_{B} e = q^2 \;\eta\; \dot c,
\label{17}
\end{eqnarray}
because the Lagrangian transforms as
\begin{equation}
\delta_{B}  L_{f} = \eta\;\frac{d}{d \tau} \;
\Bigl [ \frac{c\; ( p^2 + q^2 m^2)}{( 1 + q^2 )} + q^2\;b\;\dot c \Bigr ],
\label{18}
\end{equation}
where $\eta$ is a $\tau$-independent and a $q$-(anti)commutative {\it odd}
element ($ \eta^2 = 0 $) of a Grassmann algebra
(i.e., $\eta\;c=-q\;c\;\eta,\; \eta\;\bar c = -q\;\bar c\;\eta $)
and it commutes with
all the even fields ($x_{\mu}, p_{\mu}, e, b $) of the theory
(i.e., $\eta\;x_{\mu}=x_{\mu}\;\eta\; $etc.). It is the gauge symmetry of
the first-order Lagrangian($L_{\cal F}$) that has been exploited for the
BRST transformations (17). The reparametrization symmetry, corresponding
to the one-dimensional diffeomorphism ($\tau \rightarrow \tau
- \epsilon (\tau)$), can also be exploited for the BRST quantization. Such
a first-order Lagrangian is
\begin{equation}
L_{B}^{r}= q^{1/2} p_{\mu} \dot x^{\mu} - \; \frac{e}{1+ q^2} (p^2-m^2)
+ {\cal B}\;\dot e +\frac{{\cal B}^2}{1+q^2} + \dot {\bar \lambda}
\frac{d}{d\tau} (\lambda\;e),
\label{19}
\end{equation}
where $\cal B$ is the Nakanishi-Lautrup auxiliary field obeying the
same $q$-commutation relations as $b$ in (2) and
$\bar \lambda (\lambda)$ are the (anti)ghost fields corresponding to the
diffeomorphism transformations. These (anti)ghost fields are odd elements
of a Grassmann algebra ($\bar \lambda^2= \lambda^2 = 0,
\lambda\;\bar \lambda = - \bar \lambda\;\lambda$) and they commute
with all
the {\it even} elements of a Grassmann algebra. It can be checked that
under the
following nilpotent BRST transformations
\begin{eqnarray}
\delta_{B}^r x^{\mu}&=& \eta\;\lambda\;\dot x_{\mu}, \qquad\;
\delta_{B}^r p^{\mu} = \eta\;\lambda\;\dot p_{\mu},\;
\qquad \;\delta_{B}^r {\cal B}\; = 0, \nonumber\\
\delta_{B}^r\; e &=& \eta\; \frac{d}{d\tau} (\lambda\;e),\qquad
\delta_{B}^r \bar \lambda\;=  \eta\;{\cal B} ,
\qquad \delta_{B}^r \lambda\;
= \;\eta \lambda \dot \lambda,
\label{20}
\end{eqnarray}
the Lagrangian (19) transforms as
\begin{equation}
\delta_{B}^r  L_{B}^r = \eta\;\frac{d}{d \tau} \;
\Bigl [q^{1/2}\lambda \;p_{\mu} \dot x^{\mu}
- \; \frac{\lambda e}{1+ q^2} \;(p^2-m^2)
+ {\cal B} \frac{d}{d\tau}(\lambda\;e) \Bigr ].
\label{21}
\end{equation}
The equations of motion that emerge from (19) for $ e \neq 0 $
on the mass-shell are:
\begin{eqnarray}
\dot x_{\mu} &=& q^{1/2} \; e \; p_{\mu},
\quad \frac{d^2}{d\tau^2}(\lambda\;e) = 0,\quad
\dot p_{\mu} = 0, \quad
\ddot {\bar \lambda} = 0, \nonumber\\
\dot {\cal B}\;&=&\;- \frac{q^4}{1+q^2} ( p^2 - m^2 )= 0,
\qquad \;\dot e\;= - \;{\cal B}.
\label{22}
\end{eqnarray}

The analogue of the Euler-Lagrange equations (10) and (22) can be
obtained from the least action principle in the form of the Hamilton
equations. As a bonus, we can also derive the expressions for the
conserved charges as illustrated below:
\begin{eqnarray}
\delta S \;=\; 0
&\equiv& \int d\tau \Bigl (\delta
[q^{1/2} p_{\mu} \dot x^\mu + b\dot e
+ \dot c \Pi_{c} +\dot {\bar c} \Pi_{\bar c} \nonumber\\
&-&  H ( x_{\mu},p_{\mu},e,b, c, \bar c, \dot c, \dot {\bar c})]-
\frac{d g(\tau)}{d\tau} \Bigr ),
\label{23}
\end{eqnarray}
where $S$ is the action corresponding to the Lagrangian (8),
$H$ is the most general expression for the BRST Hamiltonian function for
a $q$-deformed free relativistic particle and
the expressions
for $g(\tau)$ are:
\begin{eqnarray}
 g(\tau) &=& \frac{c\; (p^2 + q^2 m^2)}{1+ q^2} + q^2\;b\;\dot c,\nonumber\\
 g (\tau)&=& q^{1/2}\lambda p_{\mu} \dot x^{\mu}
- \; \frac{\lambda e}{1+ q^2} (p^2-m^2) + {\cal B }
\frac{d}{d\tau}(\lambda\;e),
\label{24}
\end{eqnarray}
for the (gauge) BRST transformations (17) and
the (diffeomorphism) BRST symmetry transformations (20),
respectively.
Now, using the $q$-(anti)commutation relations
$ \dot {\bar c}\;\delta \dot c = - q\;\delta {\dot c}\;\dot {\bar c} $ and
$\delta \dot x^\mu p_{\mu}
= q \;p_{\mu} \delta \dot x^\mu $, all the variations can be taken to the
left in the corresponding terms of (23).
For the validity of the following Hamilton equations
\footnote{ In the variation of
$ \delta ( \dot c\;\Pi_{c}  + \dot {\bar c} \Pi_{\bar c})$ which is
equivalent to $ \delta ( \dot {\bar c}\;\dot c  + \dot {\bar c}\;\dot c)$ ,
we have taken $\delta \dot {\bar c}\;\dot c
- q\;\delta \dot c\;\dot {\bar c}$ from the first-term and the second-term
is expressed as $ \frac{d}{d\tau}(\delta {\bar c}\; \dot c)
-q \delta {\bar c}\; \ddot c - q\;\frac{d}{d\tau}(\delta c\; \dot {\bar c})
+ q\;\delta c\; \ddot {\bar c}$ to yield the equations of motion
$\ddot c = \ddot {\bar c} = 0$. In analogy with equations (23),(24) and (25),
it is straightforward to derive the Hamilton equations corresponding
to (22).}
\begin{eqnarray}
\dot x^\mu &=& q^{-1/2} \frac{\partial  H}{\partial p^\mu}, \quad
\dot p^\mu = - \;q^{1/2} \frac{\partial H}{\partial x^\mu}, \quad
\dot e =  \frac{\partial  H  }{\partial b},  \qquad \;
\dot b =  -\;\frac{\partial  H}{\partial e} ,\nonumber\\
\dot c &=& \frac{\partial  H  }{\partial \dot {\bar c}},\qquad \;
\ddot c = -\frac{\partial  H  }{\partial \bar c},\qquad \;
\dot {\bar c}=-q^{-1} \frac{\partial  H  }{\partial \dot c},\qquad \;
\ddot {\bar c} = q^{-1} \frac{\partial  H  }{\partial  c},
\label{25}
\end{eqnarray}
we obtain the most general expression for the conserved charge ($Q$) as:
\begin{equation}
Q =  q^{-1/2} \delta x^\mu p_{\mu} + b\;\delta e + \delta \bar c\; \dot c
-q\; \delta c \; \dot {\bar c} - g(\tau).
\label{26}
\end{equation}
The Hamilton equations of motion (25)
with the BRST Hamiltonian
\begin{equation}
 H = \frac{e}{1+ q^2} (p^2 - m^2)- \frac{b^2}{1+q^2}
+ \dot {\bar c}\; \dot c ,
\label{27}
\end{equation}
turn out to be consistent with the
Euler-Lagrange equations (10) and the contravariant metric (9).
For the global version of the BRST
symmetry transformations (17) and (20), equation (26) yields
the following charges
\begin{equation}
Q_{B} = \frac{q^2\;c\; (p^2 - m^2)}{1 + q^2} + q^2\;b\dot c
\quad \mbox{and} \quad
Q_{r} = \frac{\lambda e\;(p^2-m^2)}{1+q^2}
+ {\cal B} \frac{d}{d\tau}(\lambda\;e),
\label{28}
\end{equation}
which are found to be equivalent under the identifications $b={\cal B}$
and $ c = \lambda e$ {\it only} for $ q=\pm 1$. In fact, this
requirement ($q=\pm 1$) for the above equivalence is a manifestation of
the on-shell equivalence of the gauge and reparametrization symmetries
in the case of the deformed Lagrangian ($L_{\cal F}$).
It is interesting to check that $ (\dot c, e )\;p_{\mu}
= q\;p_{\mu}\;(\dot c, e )$, (10) and (22) lead to:
\begin{equation}
\dot Q_{B} = \frac{q^2\;\dot c\; (p^2 - m^2)}{1 + q^2} (1 - q^2)
\quad \mbox{and} \quad
\dot Q_{r} = \frac{d}{d\tau}(\lambda\;e)\;\frac{p^2-m^2}{1+q^2}
 (1-q^2).
\label{29}
\end{equation}
To have an analogy with the undeformed case ($q=1$), where the BRST charge
is conserved on any arbitrary (unconstrained) manifold, it is essential
that the deformation parameter ($q$) must be $\pm 1$ for the conservation
of the above $q$-deformed BRST charge (28). However, even for
an arbitrary value of  $q$, the BRST charge
(28) is conserved on the constrained submanifold where $p^2-m^2 = 0$.

To obtain the BRST quantization scheme, the dynamical
variables are first changed to the hermitian operators and then we
require that the physical Hilbert space must be
annihilated by the BRST operator. This, in
turn, implies that the constraint operators should annihilate the physical
states. In the $q$-deformed BRST approach, it is essential to invoke various
consistency conditions e.g. hermitian properties and the BRST algebra
to obtain a precise expression for the $q$-(anti)commutators. To
illustrate this point, we first demonstrate the correctness of $\dot Q_{B}$
of (29) in terms of the $q$-analogue of the Heisenberg equations, namely;
\begin{equation}
\dot Q_{B}\; =\; -\frac{i}{\hbar}\; [\; Q_{B}, H\; ]_{q},
\label{30}
\end{equation}
where first we define the $q$-commutators
$ [ A, B ]_{q} =  A\;B - f(q) B\; A $ in terms of an
arbitrary $q$-dependent function $f(q)$
($ f(q) \rightarrow 1 $  when $ q \rightarrow 1 $ {\it or} $ A=B $)
and do the ordering by exploiting
$q$-(anti)commutation relations (2) to obtain the desired
$q$-(anti)commutators.
For instance, using equations (27) and (28), we obtain
\begin{equation}
\dot Q_{B} = -\frac{i}{\hbar} \frac{q^2}{1+q^2}
\Bigl (\; [\; c\;(p^2-m^2), \dot {\bar c}\; \dot c\; ]_{q}
+ [\; b \;\dot c,  e\;(p^2 - m^2)\; ]_{q}\; \Bigr ).
\label{31}
\end{equation}
Now, using the above definition of the $q$-commutator and exploiting the
relations $ \dot c\;(p_{\mu}, m) = q\; (p_{\mu},m)\; \dot c,\;
\dot {\bar c}\; (p_{\mu},m) = q\; (p_{\mu},m) \;\dot {\bar c} $ and
$ \dot c\; c = -\frac{1}{q}\; c \;\dot c $, the first $q$-commutator
in (31) can be
converted into a $q$-anticommutator and the second commutator can be
reordered
using $ b\;(p_{\mu},m)= q \;(p_{\mu},m)\; b,\; \dot c\;(p_{\mu}, m)
= q\; (p_{\mu},m)\; \dot c,\; $ and
$ \dot c \;e = e \;\dot c $ to yield the right hand side of (31) as
\begin{equation}
-\frac{i}{\hbar} \frac{q^2}{1+q^2}\;
\Bigl [\; \frac {\{ c, \;\dot {\bar c} \}_{q}}{q^4} + [\;b,e\;]_{q} \;
\Bigr ]\;
\dot c\; (p^2 - m^2),
\label{32}
\end{equation}
where the $q$-(anti)commutators are
\begin{equation}
 \{ c, \dot {\bar c} \}_{q}
=   c \;\;\dot {\bar c} + q^3\;f(q)\; \dot {\bar c}\; c ,\qquad
[ b,\; e ]_{q} = b\; e - \frac{g(q)}{q^4}\; e\; b.
\label{33}
\end{equation}
Here arbitrary $q$-dependent functions
$g(q)$ and $f(q)$ reduce to one as $ q \rightarrow 1 $.
Comparison and consistency with (29) yields {\it one} of the solutions as:
\begin{equation}
\{ c, \dot {\bar c} \}_{q} \; = i\; \hbar\;q^4, \; \qquad \;
[ b, e ]_{q} = - i\;\hbar\;q^2.
\label{34}
\end{equation}
The hermiticity requirement on the above $q$-(anti)commutators
leads to
\begin{eqnarray}
\Bigl ( |q|^6\;|f(q)|^6 - 1 \Bigr )\;\dot {\bar c}\;c &=&
i\;\hbar\;{q^{*}}^3\; \Bigl ( q^4\;f^{*}(q) - q^{*} \Bigr ),\nonumber\\
\Bigl ( |q|^8 - |g(q)|^2 \Bigr )\;e\;b &=& i\;\hbar\;q^4\;
\Bigl ( {q^{*}}^6
- q^2\;g^{*}(q) \Bigr ),
\label{35}
\end{eqnarray}
as the general restriction on $g(q)$ and $f(q)$ (see, e.g.,
Aref'eva and Volovich Ref.[6]). One of the trivial solutions
($ g(q) = q^4,\; f(q) = q^{-3},\; q^2 = {q^{*}}^2,\; q^4 = {q^{*}}^4)$
implies that $q^2$ and $q^4$ are real parameters. Under such restrictions,
the (anti)commutators (34)
\begin{equation}
\{ c, \dot {\bar c} \}_{q} \equiv c\;\dot {\bar c}
+ \dot {\bar c}\; c
= i\; \hbar\;q^4, \;\qquad\;
[ b, e ]_{q} \equiv b\;e - e\;b = - i\;\hbar\;q^2,
\label{36}
\end{equation}
reduce to the corresponding undeformed BRST (anti)commutators for
$q=\pm 1$. The nilpotency of the
$q$-BRST charge $Q_{B}^2 = \frac{1}{2} \{ Q_{B}, Q_{B} \}_{q} = 0$ is
trivially satisfied because of the absence of the canonically
conjugate variables in the expression for $Q_{B}$. To complete the BRST
algebra, we further require the
validity of the relation
($-\frac{i}{\hbar}[  Q_{c}, Q_{B} ]_{q} = Q_{B}$)
where the ghost charge $ Q_{c}= c\;\dot {\bar c} + \bar c\; \dot c $,
emerging due to the global scale invariance, is conserved
only for $q=1$ (and $[Q_{c},Q_{c}]_{q}=0$). This $q$-commutator is
succinctly expressed as:
\begin{equation}
-\frac{i}{\hbar}\;[  Q_{c}, Q_{B} ]_{q}
= -\frac{i}{\hbar}\;\frac{q^2}{1+q^2} [ c\;\dot {\bar c}, c\;(p^2-m^2) ]_{q}
-\frac{i}{\hbar}\;q^2 [ \bar c\;\dot c, b\;\dot c ]_{q}.
\label{37}
\end{equation}
In the computation of the first $q$-commutator in (37), we choose the
arbitrary function such that we are consistent with the
$q$-anticommutator (36). For instance, after reordering, this $q$-commutator
becomes
$-\frac{i}{\hbar}\;\frac{q^2}{1+q^2} \; c ( \dot {\bar c}\;c
+ \frac{F(q)}{q^4}\; c\;\dot {\bar c}\;) (p^2 - m^2) $. Now, choosing
$ F(q) = q^4 $, we obtain this $q$-commutator
as $ \frac{q^2}{1+q^2}\; q^4\; c\;(p^2 - m^2) $.
We exploit an analogous procedure for the computation of the second
$q$-commutator which ultimately reduces to
$ \frac{i}{\hbar}\;q^2\;b\; \{ \bar c, \dot c \}_{q}\;\dot c $ where
$\{ \bar c, \dot c \}_{q}= \bar c\;\dot c + G(q) \dot c\;\bar c $ with an
arbitrary function $G(q)$. For the sum of these two $q$-commutators to
yield $Q_{B}$, we require:
\begin{equation}
\{ \bar c, \dot c \}_{q}=- i \hbar, \qquad  q^4 = 1.
\label{38}
\end{equation}
The hermiticity requirement on the above $q$-anticommutator implies $G(q)=1$.
A definite and  sensible expression for $Q_{B}$, however, requires that $q$
must be $\pm 1$.

To compute the $q$-commutator between $x_{\mu}$ and $p_{\nu}$, we first
define a relationship between the basic $q$-commutator
$[ x_{\mu}, p_{\nu} ]_{q}$ and a $q$-Poisson bracket
$\{ x_{\mu}, p_{\nu}\}_{q}^{PB}$ as
\begin{equation}
[\; x_{\mu}, p_{\nu}\; ]_{q} =  i\;\hbar\; M(q)\;
\{ x_{\mu}, p_{\nu}\}_{q}^{PB},
\label{39}
\end{equation}
where $\{ x_{\mu}, p_{\nu}\}_{q}^{PB}= q^{-1/2} \; \eta_{\mu\nu}$
due to symplectic metric (9) and
$[ x_{\mu}, p_{\nu} ]_{q} = x_{\mu}\; p_{\nu} - N(q)\; p_{\nu} \; x_{\mu}$.
Here $q$-dependent functions $M(q)$ and $N(q)$ go to one as
$q \rightarrow 1$. The hermitian condition on (39) yields
one of the solutions as:
\begin{equation}
|N(q)|^2\; =\; 1, \quad \mbox{and} \quad M(q)\;q^{-1/2}\;
= \;\frac{M^{*}(q)}{N^{*}(q)} \;\; {q^{*}}^{-1/2}.
\label{40}
\end{equation}
With the basic definition (39), we obtain
\begin{equation}
[ x_{\mu}, p^2 ]_{q} = x_{\mu}\; p^2 - N^2(q)\; p^2 \; x_{\mu}
\equiv i\;\hbar\; M(q)\; \{ x_{\mu}, p^2 \}_{q}^{PB},
\label{41}
\end{equation}
where $\{ x_{\mu}, p^2 \}_{q}^{PB}= q^{-1/2}\;(1 + q^2)\;p_{\mu}$ fixes $N(q)$
to be $q^2$ and, therefore, $|q|^4=1 $. Now, we require the validity of
equation (10) by exploiting (anti)commutators (36), (38) and (41)
in the Heisenberg equations of motion. For instance, the "time" derivative of
$x_{\mu}$ can be expressed in terms of the $q$-BRST Hamiltonian $H$  as:
\begin{equation}
\dot x_{\mu} = - \frac{i} {\hbar} [ x_{\mu}, H ]_{q} \equiv
q^{1/2}\; e\;p_{\mu}.
\label{42}
\end{equation}
In the computation of
$ [ x_{\mu}, e\;p^2 ]_{q} = x_{\mu}\;e\;p^2 - h(q)\;e\;p^2\;x_{\mu} $, we
do the reordering using $ e\;x_{\mu} = q\;x_{\mu}\;e $ and require the
consistency with (41) which fixes $h(q)= q^3$. Finally equality in
(42) leads to $M(q) = q^2$. This, in turn, yields $q= \pm 1$ due to the
requirement (40) (for the real value of $q$). Similarly, rest of
the equations of motion (10) can be checked to be satisfied only for
$q= \pm 1$ if we use the $q$-(anti)commutators (36) and (38) in the
Heisenberg equations of motion.

The key ingredients in our quantization scheme are hermitian condition
on $q$-(anti)commutators, validity of the $q$-BRST algebra, conservation
of the BRST charge on an unconstrained manifold and the requirement that
the on-shell condition should remain intact under $q$-deformed Heisenberg
equations of motion. In the limit when $ q\rightarrow 1,
\hbar \rightarrow 0 $, we obtain classical relations and in the limit
$ q\rightarrow 1$ the usual quantum mechanical (anti)commutators emerge
automatically. We hope the  $q$-deformed Hamiltonian
formulation for this system with $q$-deformed
Dirac brackets, $q$-deformed constraint analysis, etc.,
would be able to shed more light on the
quantization scheme for any arbitrary value of $q$ [11].

{\it Fruitful conversations with A.Filippov and P.Pyatov are gratefully
acknowledged. Thanks are also due to Ann for carefully reading the
manuscript.}


\begin{thebibliography}{99}
\bibitem{1}  V.G.Drinfeld, {\it Quantum Groups }, Proc.Int.Cogr.
             Math., Berkeley, { \bf 1} (1986) 798.
             M.Jimbo,  Lett.Math.Phys. {\bf 10} (1985) 63, {\bf 11}
             (1986) 247.
\bibitem{2}  S.L.Woronowicz, Comm.Math.Phys. {\bf 111} (1987) 613, {\bf 122}
             (1989) 125.
\bibitem{3}  See, e.g., for review, S.Majid, Int. J. Mod. Phys. {\bf A5}
             (1990) 1.
\bibitem{4}  See, e.g., A.Schirrmacher, J.Wess and B.Zumino,
             Z.Phys. {\bf C49} (1991) 317.
\bibitem{5}  I.Ya.Aref'eva and I.V.Volovich, Mod.Phys.Lett. {\bf A6}
             (1991) 893, Phys.Lett. {\bf 264B} (1991) 62,
             A.P.Isaev and Z.Popowicz,  Phys.Lett. {\bf 307B} (1993) 353.
\bibitem{6}  I.Ya.Aref'eva and I.V.Volovich,
             Phys.Lett. {\bf 268B} (1991) 179,\\ R.M.Mir-Kasimov, J.Phys.{\bf
             A} {\bf 24} (1991) 4283, J.Schwenk and J.Wess,
             Phys.Lett. {\bf 291B} (1992) 273, V.Spiridonov, Phys.Rev.Lett.
             {\bf 69} (1992) 398, A.P.Isaev and R.P.Malik,
             Phys.Lett. {\bf 280B} (1992) 219.
\bibitem{7}  H.S.Snyder, Phys.Rev. {\bf 71} (1947) 38,
             C.N.Yang, Phys.Rev. {\bf 72} (1947) 874,
             H.Yukawa, Phys.Rev. {\bf 91} (1953) 415.
\bibitem{8}  I.V.Volovich, Class.Quant.Grav. {\bf 4} (1987) L83,
             CERN preprint, CERN-TH-4781 (1987).
\bibitem{9}  M.N.Barber and P.A.Pearce (Eds.),
             {\it Yang-Baxter Equations, Conformal Invariance and
             Integrability in Statistical Mechanics and Field Theory}
             Int. J.Mod.Phys. {\bf B4,5} (1990),
             C.N.Yang and M.L.Ge (Eds.),
             {\it Braid Groups, Knot Theory and Statistical Mechanics }
             World Scientific, Singapore (1989),
             G.Moore and N.Seiberg, Comm.Math.Phys. {\bf 123} (1989) 177,
             D.Bonatsos, E.N.Argyres, S.B.Drenska, P.P.Raychev, R.P.Roussev
             and Yu.F.Smirnov, Phys.Lett. {\bf 251B} (1990) 477.
\bibitem{10} R.P.Malik, Phys.Lett. {\bf 316B} (1993) 257,
             Phys.Lett. {\bf 345B} (1995) 131.
\bibitem{11} R.P.Malik,  (in preparation).
\bibitem{12} L.Brink, S.Deser, B.Zumino, P.Di Vecchia and P.Howe,
             Phys.Lett. {\bf 64B} (1976) 435, L.Brink, P.Di Vecchia and
             P.Howe, Nucl.Phys. {\bf B118} (1977) 76.
\bibitem{13} C.Becchi, A.Rouet and R.Stora, Phys.Lett.{\bf 52B} (1974) 344,
             I.V.Tyutin, Lebedev Report No. {\bf 39} (1975) (Unpublished).
\bibitem{14} See, e.g., D.Nemeschansky, C.Preitschopf and M.Weinstein,
             Ann.Phys. {\bf 183} (1988) 226.
\bibitem{15} J.Wess and B.Zumino, Nucl.Phys. (Proc.Suppl)
             {\bf B18} (1990) 302.
\end{thebibliography}
\end{document}